\documentclass[preprint,preprintnumbers,amsmath,amssymb,double-spaced]{revtex4}
\usepackage{graphicx}
\usepackage{epstopdf}
\usepackage{dcolumn}
\usepackage{bm}

\begin{document}

\title{Thermal Quantum Speed Limit for Classical-Driving Open Systems}

\author{Wenjiong Wu}

\author{Kai Yan}

\author{Xiang Hao}
\altaffiliation{Corresponding author}
\email{haoxiang_edu198126@163.com}

\affiliation{School of Mathematics and
Physics, Suzhou University of Science and Technology, Suzhou,
Jiangsu 215011, People's Republic of China}

\begin{abstract}

 Quantum speed limit ($\mathrm{QSL}$) time for open systems driven by classical fields is studied in the presence of thermal bosonic environments. The decoherence process is quantitatively described by the time-convolutionless master equation. The evolution speed of an open system is related not only to the strength of driving classical field but also to the environmental temperature. The energy-state population plays a key role in the thermal $\mathrm{QSL}$. Comparing with the zero-temperature reservoir, we predict that the structural reservoir at low temperature may contribute to the acceleration of quantum evolution. The manifest oscillation of $\mathrm{QSL}$ time takes on under the circumstance of classical driving field. We also investigate the scaling property of $\mathrm{QSL}$ time for multi-particle noninteracting entangled systems. It is demonstrated that entanglement of open systems can be considered as one resource for improving the potential capacity of thermal quantum speedup.

\vspace{1.6cm} PACS numbers: 03.65.-w, 03.65.Yz, 03.67.Lx, 42.50.-p

\end{abstract}

\maketitle

\section{Introduction}

As progress of quantum theory \cite{Lloyd00,Giovanetti11,Caneva09,Deffner10,Mukherjee13,Deffner14,Chin12,Hegerfeldt14,Tsang13,Giovannetti11} moves far ahead, the speed of a quantum evolution gradually becomes a hot issue in the fields of quantum communication and quantum metrology. The $\mathrm{QSL}$ time \cite{Taddei13,Deffner13,Campo13,Jones10,Fung14,Khalil15} is referred to as the minimal evolution time of an arbitrarily driven open quantum system \cite{Xu2014,Obada11,Carlini08,Latune13}. To distinguish an initial state and a final state, a geometric approach is provided by the Bures angle $B(\rho_0,\rho_t)$ \cite{Deffner13,Brody03}. This feasible measure can deduce two types of $\mathrm{QSL}$ time, $\tau_{QSL}$, i.e., Mandelstam-Tamm bound (MT) \cite{Mandelstam1945} and Margolus-Levitin bound (ML) \cite{Margolus98}. Very recently, some researchers have put forward other methods including fidelity approach \cite{Zhang14}, metric approach \cite{Xu15}, one based on quantum Fisher information \cite{Frowis12}. With respect to the above measurements, $\tau_{QSL}$ is closely dependent on an actual driving time $\tau_D$ for an open system. As noted in Refs. \cite{Xu14,Liu15}, in the situation $\frac{\tau_{QSL}}{\tau_D}=1$, it indicates that the quantum evolution has no potential capacity for further acceleration. However, in the situation $\frac{\tau_{QSL}}{\tau_D}<1$, the smaller ratio $\tau_{QSL}/{\tau_D}$ the greater this potential capacity for speedup will be. How to obtain a smaller quantum speed limit time with respect to a given driving time is a valuable question in the field of open quantum information processing. In response to this issue, several feasible solutions are proposed such as changing environmental factors \cite{Deffner13,Zhang15} or initial state \cite{Frowis12,Liu15}. For example, we can acquire acceleration of quantum evolution by strengthening the system-environment couplings \cite{Deffner13} or selecting some kinds of special state \cite{Liu15,Wu15} as an initial state. For explaining the mechanism for the speedup, there is one viewpoint which is focused on the effects of non-Markovianity \cite{Laine12,Rivas10,Li10,Ma14} for a dynamical map. According to some previous results, entanglement is thought of to be possible resources for the acceleration of the evolution \cite{Giovannetti03,Batle05,Borras06}.

Nowadays, the control of quantum systems by electromagnetic fields plays a significant role in a variety of applications of physics.  As we know, the weak interaction between systems and environments is always unavoidable in real experiments. However, in this weak-coupling limit, it is difficult to engender the potential capacity to accelerate the evolution without any external operation. This motivates us to find out what kind of external
conditions can be manipulated in order to induce the occurrence of quantum speedup potential in practical operations. To reasonably consider the effects of external driving and environment temperatures, we apply the second-order time-convolutionless master equation to describe the dynamical behavior of the system. In ultra-low temperature condition, the single-excitation approximation used by \cite{Deffner13} is not enough. We make a weak-coupling approximation in order to use the Lorentzian spectral density approach. In the following context, we will prove that high-excitation states of the system result in some high-frequency oscillating terms which can be neglected in the weak-coupling limit. Besides it, we also consider another equivalent dissipation model to demonstrate the validity of weak-coupling approximation.

In this paper, we consider noninteracting two-level atoms driven by classical laser fields, which are respectively coupled to independent finite-temperature leaky cavities. The paper is organized as follows. In Sec. II, we introduce the model of a driven two-level system interacting with a thermal bosonic environment in weak coupling regime. The expression of the reduced density matrix for single system is obtained by the time-convolutionless quantum master equation under weak coupling approximation. The technique of operator projection for open dynamics is then used to derive a general evolution of $N$-particle open systems. In Sec. III, we define a $\tau_{QSL}$ measure based on trace distance. It is found out that the temperature of the reservoir and the driving strength of the laser field may give rise to the acceleration of quantum decoherence with respect to an excited state of single system. For a given driving time, the change of excited-state population dominates in the measure for $\mathrm{QSL}$ time. With the consideration of multi-qubit entanglement of open systems, we study the scaling property of $\mathrm{QSL}$ time. Finally, a simple discussion concludes the paper.

\section{The Model}

In many applications of quantum optics, a two-level atom coupled to a structural reservoir at finite temperature is referred to as a typical model. Hereby we make use of classical lasers to drive $N$ identical atoms independently coupled to their respective leaky cavities. The total Hamiltonian of the system and environment is given by
\begin{equation}
\label{eq:1}
\hat{H}^{(tot)}= \hat{H}_{S}^{(tot)}+\hat{H}_{E}^{(tot)}+\hat{H}_{I}^{(tot)}.
\end{equation}
The Hamiltonian of the whole driven system can be written as
\begin{equation}
\label{eq:2}
\hat{H}_{S}^{(tot)}= \sum_{k=1}^N\frac{\omega_0}{2}\hat{\sigma}_z^{(k)}+\Omega(e^{-i\omega_Lt}\hat{\sigma}_+^{(k)}+e^{i\omega_Lt}\hat{\sigma}_-^{(k)}),
\end{equation}
where $\hat{\sigma}_+^{(k)}=|e\rangle_{k}\langle{g}|, \hat{\sigma}_-^{(k)}=|g\rangle_{k}\langle{e}|, \hat{\sigma}_z^{(k)}=|e\rangle_{k}\langle{e}|-|g\rangle_{k}\langle{g}|$. $\omega_0$ denotes the energy transition between an excited state $|e\rangle_{k}$ and a ground state $|g\rangle_{k}$. $\omega_L$ is the frequency of driving laser field. $\Omega$ represents the coupling strength between each atom and laser field. The Hamiltonian of these independent thermal environments is obtained as
\begin{equation}
\label{eq:3}
\hat{H}_{E}^{(tot)}=\sum_{k=1}^{N}\sum_{j}\omega_{j}\hat{b}^{\dag(k)}_{j}\hat{b}^{(k)}_{j},
\end{equation}
where $\hat{b}^{(k)}_{j}$ and $\hat{b}^{\dag(k)}_{j}$ are the annihilation and creation operator in the $k-$th bosonic environment Hilbert space. The interaction term can be expressed as
\begin{equation}
\label{eq:4}
\hat{H}_{I}^{(tot)}=\sum_{k=1}^{N}\sum_j(g_j\hat{b}_j^{\dag(k)}\hat{\sigma}_-^{(k)}+g_j^*\hat{b}_j^{(k)}\hat{\sigma}_+^{(k)}),
\end{equation}
where $g_j$ represents the coupling strength of the interactions of each atom with the $j-$th mode of the corresponding field.
By performing the rotating unitary operation $\hat{U}_1=\prod_{k=1}^{\otimes N} \; \; e^{\frac{1}{2}(-i\omega_Lt \; \hat{\sigma}_z^{(k)})}$, the total Hamiltonian
is transformed to an effective Hamiltonian,
\begin{eqnarray}
\label{eq:5}
\hat{H}&=&\sum_{k=1}^{N}\frac{1}{2}[(\omega_0-\omega_L)\hat{\sigma}_z^{(k)}+\Omega\hat{\sigma}_x^{(k)}] \nonumber\\
&+&\sum_{k=1}^{N}\sum_{j}(g_{j}e^{-i\omega_Lt}\hat{b}^{\dag(k)}_{j}\hat{\sigma}_{-}^{(k)}+g^{\ast}_{j}e^{i\omega_Lt}\hat{b}^{(k)}_{j}\hat{\sigma}^{(k)}_{+})
+\sum_{k=1}^{N}\sum_{j}\omega_{j}\hat{b}^{\dag(k)}_{j}\hat{b}^{(k)}_{j},
\end{eqnarray}
where $\hat{\sigma}^{(k)}_x=\frac{1}{2}(\hat{\sigma}_+^{(k)}+\hat{\sigma}^{(k)}_-)$.

For the $k-$th atom, in the interaction representation, the decoherence of the time-dependent state $\hat{\rho}^{(k)}_t$ can approximately satisfy the second-order time-convolutionless master equation,
\begin{equation}
\label{eq:6}
\frac {d\hat{\rho}^{(k)}_t}{dt}=-\int_{0}^{t}dt_{1} \mathrm{Tr}_{E}[\hat{H}^{\prime(k)}_{I}(t),[\hat{H}^{\prime(k)}_{I}(t_1),\hat{\rho}^{(k)}_t\otimes\hat{\rho}^{(k)}_{E}]],
\end{equation}
where $\hat{H}^{\prime(k)}_{I}(t)=e^{it(\hat{H}^{(k)}_S+\hat{H}^{(k)}_{E})}\hat{H}^{(k)}_{I}e^{-it(\hat{H}^{(k)}_S+\hat{H}^{(k)}_{E})}$, $\hat{H}^{(k)}_S=\frac{1}{2}[(\omega_0-\omega_L)\hat{\sigma}_z^{(k)}+\Omega\hat{\sigma}_x^{(k)}]$ represents the Hamiltonian of the $k-th$ driven system,
$\hat{H}^{(k)}_{E}=\sum_{j}\omega_{j}\hat{b}^{\dag(k)}_{j}\hat{b}^{(k)}_{j}$ represents the Hamiltonian of the $k-th$ independent thermal environment and $\hat{H}^{(k)}_{I}=\sum_{j}(g_{j}e^{-i\omega_Lt}\hat{b}^{\dag(k)}_{j}\hat{\sigma}_{-}^{(k)}+g^{\ast}_{j}e^{i\omega_Lt}\hat{b}^{(k)}_{j}\hat{\sigma}^{(k)}_{+})$
the Hamiltonian of interaction term. The notation $\mathrm{Tr}_{E}$ is the partial trace over the freedom of the $k-th$ environment. $\hat{\rho}^{(k)}_{E}=\exp(-\hat{H}^{(k)}_{E}/\kappa_{B}T)/\mathrm{Tr}[\exp(-\hat{H}^{(k)}_{E}/\kappa_{B}T)]$ is the thermal equilibrium state of the $k-th$ environment and follows $\mathrm{Tr}[\hat{H}^{\prime}_{I}(t)\hat{\rho}^{(k)}_{E}]=0$ \cite{Goan11}. Here, it is assumed that the Boltzmann constant $\kappa_{B}$ and Planck constant $h$ are $1$ and the dimensionless low temperature condition of $T \ll \omega_0$ is considered.

In the diagonalization representation of each driven atom, the effective Hamiltonian for the $k-$th system can be written as $\hat{\bar{H}}^{(k)}_{S}=\frac{\omega_s}{2}\hat{\tau}_z^{(k)}$ where $\omega_s=\sqrt{(\omega_0-\omega_L)^2+\Omega^2}$. The effective operator is constructed by $\hat{\tau}_z^{(k)}=|1\rangle_k\langle 1|-|0\rangle_k\langle 0|$ where $|1 \; (0) \rangle_k$ are the eigenstates of the Hamiltonian of each driven atom. The interaction Hamiltonian in this dressed state basis $|1 \; (0) \rangle_k$ can be given by
\begin{equation}
\label{eq:7}
\hat{\bar{H}}^{\prime(k)}_{I}(t)=\hat{A}^{\dag}(t)\otimes \hat{B}(t)+\hat{A}(t)\otimes \hat{B}^{\dag}(t).
\end{equation}
Here $\hat{A}(t)=\sum_{j}g_j e^{-i\omega_j t}\hat{b_j}$ and $\hat{B}^{\dag}(t)=C_0 \hat{\tau}_z^{(k)}+C_{+} e^{i\omega_s t}\hat{\tau}_+^{(k)}-C_{-} e^{-i\omega_s t}\hat{\tau}_-^{(k)}$ where $C_\pm=\frac{\omega_s\pm(\omega_0-\omega_L)}{2\omega_s}$ and $C_0=\sqrt{C_+C_-}$.
Then the expression of the time-convolutionless master equation in the dressed state basis is given as \cite{Hao2013,Hao13}
\begin{equation}
\label{eq:8}
\frac {d\hat{\bar{\rho}}^{(k)}_t}{dt}=-i[\hat{\bar{H}}^{(k)}_{S}+\hat{\bar{H}}^{\prime},\hat{\bar{\rho}}^{(k)}_t]+\sum_{m=\pm,z}\gamma_m(t)\; [\hat{\tau}_m^{(k)}\hat{\bar{\rho}}^{(k)}_t\hat{\tau}^{\dag(k)}_m-\frac 12\{\hat{\tau}^{\dag(k)}_m\hat{\tau}_m^{(k)}, \hat{\bar{\rho}}^{(k)}_t \}]
+\hat{O}[\hat{\bar{\rho}}^{(k)}_t],
\end{equation}
where $\gamma_m(t)(m=\pm,z)$ is the time-dependent decay rates and
they are calculated as $\gamma_{+}(t)=2C^{2}_+\mathrm{Re}(\Gamma_{+})+2C^{2}_-\mathrm{Re}(\Gamma^{\prime}_{-})$,
$\gamma_{-}(t)=2C^{2}_-\mathrm{Re}(\Gamma_{-})+2C^{2}_+\mathrm{Re}(\Gamma^{\prime}_{+})$ and
$\gamma_{z}(t)=2C^2_0\mathrm{Re}(\Gamma_0+\Gamma^{\prime}_0)$. $\hat{\bar{H}}^{\prime}=\mathrm{Im}(\Gamma_0-\Gamma^{\prime}_0)C^2_{0}\hat{\tau}_z^{2(k)}+\sum_{q=\pm}\mathrm{Im}(\Gamma_{q}-\Gamma^{\prime}_{q})C^2_{q}\hat{\tau}^{\dag(k)}_{q}\hat{\tau}^{(k)}_{q}$. The notation $\mathrm{Im(Re)}$ denotes imaginary (real) part of a complex parameter. The parameters $\Gamma_{q}$ and $\Gamma^{\prime}_{q}(q=0,\pm)$ are determined by
\begin{eqnarray}
\label{eq:9}
\Gamma_{q}&=&\int_{0}^{t}dt_{1}\sum_{j}|g_j|^2\cdot \bar{n}_{j}e^{i(\omega_j-\omega_L-q\omega_s)(t-t_1)} \nonumber \\
\Gamma^{\prime}_{q}&=&\int_{0}^{t}dt_{1}\sum_{j}|g_j|^2\cdot(\bar{n}_{j}+1) e^{i(\omega_j-\omega_L-q\omega_s)(t-t_1)},
\end{eqnarray}
where $\bar{n}_j=(e^{\omega_j/T}-1)^{-1}$ is the mean number for the $j$th mode of the thermal environment at $T$ temperature.
The strength of the couplings can be described by the spectral
function as $J(\omega)=\sum_j | g_j|^{2}\delta(\omega-\omega_{j})$.
The last term in Eq. (8) is
\begin{widetext}
\begin{eqnarray}
\label{eq:10}
\hat{O}[\hat{\bar{\rho}}^{(k)}_t]&=&\Gamma_{0}\cdot[C_0C_{+}(\hat{\tau}_z^{(k)}\hat{\bar{\rho}}^{(k)}_t\hat{\tau}_-^{(k)}-\hat{\bar{\rho}}^{(k)}_t\hat{\tau}_-^{(k)}\hat{\tau}_z^{(k)})-C_0C_{-}(\hat{\tau}_z^{(k)}\hat{\bar{\rho}}^{(k)}_t\hat{\tau}_+^{(k)}-\hat{\bar{\rho}}^{(k)}_t\hat{\tau}_+^{(k)}\hat{\tau}_z^{(k)})]\nonumber \\
&+&\Gamma_{+}\cdot[C_0C_{+}(\hat{\tau}_+^{(k)}\hat{\bar{\rho}}^{(k)}_t\hat{\tau}_z^{(k)}-\hat{\bar{\rho}}^{(k)}_t\hat{\tau}_z^{(k)}\hat{\tau}_+^{(k)})-C_{+}C_{-}(\hat{\tau}_+^{(k)}\hat{\bar{\rho}}^{(k)}_t\hat{\tau}_+^{(k)}-\hat{\bar{\rho}}^{(k)}_t\hat{\tau}_+^{(k)}\hat{\tau}_+^{(k)})]\nonumber\\
&-&\Gamma_{-}\cdot[C_0C_{-}(\hat{\tau}_-^{(k)}\hat{\bar{\rho}}^{(k)}_t\hat{\tau}_z^{(k)}-\hat{\bar{\rho}}^{(k)}_t\hat{\tau}_z^{(k)}\hat{\tau}_-^{(k)})+C_{+}C_{-}(\hat{\tau}_-^{(k)}\hat{\bar{\rho}}^{(k)}_t\hat{\tau}_-^{(k)}-\hat{\bar{\rho}}^{(k)}_t\hat{\tau}_-^{(k)}\hat{\tau}_-^{(k)})]\nonumber\\
&+&\Gamma^{\prime}_{0}\cdot[C_0C_{+}(\hat{\tau}_-^{(k)}\hat{\bar{\rho}}^{(k)}_t\hat{\tau}_z^{(k)}-\hat{\bar{\rho}}^{(k)}_t\hat{\tau}_z^{(k)}\hat{\tau}_-^{(k)})-C_0C_{-}(\hat{\tau}_+^{(k)}\hat{\bar{\rho}}^{(k)}_t\hat{\tau}_z^{(k)}-\hat{\bar{\rho}}^{(k)}_t\hat{\tau}_z^{(k)}\hat{\tau}_+^{(k)})]\nonumber \\
&+&\Gamma^{\prime}_{+}\cdot[C_0C_{+}(\hat{\tau}_z^{(k)}\hat{\bar{\rho}}^{(k)}_t\hat{\tau}_+^{(k)}-\hat{\bar{\rho}}^{(k)}_t\hat{\tau}_+^{(k)}\hat{\tau}_z^{(k)})-C_+C_{-}(\hat{\tau}_+^{(k)}\hat{\bar{\rho}}^{(k)}_t\hat{\tau}_+^{(k)}-\hat{\bar{\rho}}^{(k)}_t\hat{\tau}_+^{(k)}\hat{\tau}_+^{(k)})]\nonumber\\
&-&\Gamma^{\prime}_{-}\cdot[C_0C_{-}(\hat{\tau}_z^{(k)}\hat{\bar{\rho}}^{(k)}_t\hat{\tau}_-^{(k)}-\hat{\bar{\rho}}^{(k)}_t\hat{\tau}_-^{(k)}\hat{\tau}_z^{(k)})+C_+C_{-}(\hat{\tau}_-^{(k)}\hat{\bar{\rho}}^{(k)}_t\hat{\tau}_-^{(k)}-\hat{\bar{\rho}}^{(k)}_t\hat{\tau}_-^{(k)}\hat{\tau}_-^{(k)})]\nonumber\\
&+&\mathrm{h.c.}
\end{eqnarray}
\end{widetext}
The notation $\mathrm{h.c.}$ represents Hermitian conjugate. The Lamb shift Hamiltonian $\hat{\bar{H}}^{\prime}$ describes a small shift in the energy of the eigenvectors of $\hat{\bar{H}}^{(k)}_{S}$ and may be neglected because it has no qualitative effect on the decoherence of the system.
The high-frequency oscillating term $\hat{O}[\hat{\bar{\rho}}^{(k)}_t]$ is neglected in the weak-coupling condition \cite{Breuer01}.

The thermal environment can be described by an effective Lorentzian spectral density of the form $J(\omega)=\frac{1}{2\pi}\frac{\gamma_0\lambda^2}{(\omega_0-\omega)^2+\lambda^2}$, where $\gamma_0$ is the coupling strength and $\lambda$ the spectral width. In particular, our issue belongs to the case of a weak-coupling regime, i.e., $\gamma_0<\frac{1}{2}\lambda$. By the technique of projection operator \cite{Bellomo07,Wang13,Hao2013}, the above dynamical map in the form of the master equation in the dressed state basis can be equivalently written by $\hat{\bar{\rho}}^{(k)}_t= \hat{\varepsilon}^{(k)}_t \; \hat{\bar{\rho}}^{(k)}(0)$. The projection operator $\hat{\varepsilon}^{(k)}_t$ is expressed as
\begin{eqnarray}
\label{eq:11}
\hat{\varepsilon}^{(k)}_t(|1\rangle\langle1|)&=&a|1\rangle_k\langle1|+(1-a)|0\rangle_k\langle0| \nonumber\\
\hat{\varepsilon}^{(k)}_t(|1\rangle\langle0|)&=&b|1\rangle_k\langle0| \nonumber\\
\hat{\varepsilon}^{(k)}_t(|0\rangle\langle1|)&=&b^{\ast}|0\rangle_k\langle1| \nonumber\\
\hat{\varepsilon}^{(k)}_t(|0\rangle\langle0|)&=&c|1\rangle_k\langle1|+(1-c)|0\rangle_k\langle0|,
\end{eqnarray}
where $a=\frac{1}{2}+\frac{1}{2}e^{-p(t)}[\delta(t)+1]$, $b=e^{-r(t)}e^{-i\omega_st}$, $c=\frac{1}{2}+\frac{1}{2}e^{-p(t)}[\delta(t)-1]$. The parameters are determined by the decay rates $r(t)=\frac 12 \int_{0}^{t}dt_{1}[\gamma_{+}(t_1)+\gamma_{-}(t_1)+4\gamma_{z}(t_1)]$, $p(t)=\int_{0}^{t}dt_{1}[\gamma_{+}(t_1)+\gamma_{-}(t_1)]$ and $\delta(t)=\int_{0}^{t}dt_{1}e^{p(t)}[\gamma_{+}(t_1)-\gamma_{-}(t_1)]$. For $N-$particle noninteracting atoms, the evolved density matrix for the whole open system can be generally expressed as
\begin{equation}
\label{eq:12}
\hat{\bar{\rho}}_t=\hat{\varepsilon}^{(N)}_t \otimes \hat{\varepsilon}^{(N-1)}_t \otimes \cdots \otimes \hat{\varepsilon}^{(1)}_t \; \hat{\bar{\rho}}(0).
\end{equation}

According to the result of \cite{Bellomo07}, we consider the whole system formed by $N$ noninteracting parts $\{k=1,2,...,N \}$, each part consisting of an atom locally interacting with a reservoir. In this assumption of independent parts, the time evolution operator factorizes as $\hat{U}^{\mathrm{tot}}(t)=\Pi_{k=1}^{\otimes N}\hat{U}^{k}(t)$. The state of each part evolves as $\hat{\bar{\rho}}^{(k)}_t=\mathrm{Tr}\{ \hat{U}^{k}(t) \hat{\bar{\rho}}^{(k)}(0)\otimes \hat{\bar{\rho}}^{(k)}_{E}(0) \hat{U}^{k \dag}(t) \} =\hat{\varepsilon}^{(k)}_t \; \hat{\bar{\rho}}^{(k)}(0)$. Therefore, if we derive the evolution of each independent part, the dynamics of the whole system can be easily obtained.

\section{Quantum speed-up dynamics}

For a driven open quantum system, the minimal time for the evolution from an initial state $\rho_0$ to
a final state $\rho_{t}$, i.e., $\tau_{QSL}$, is meaningful. We need to choose appropriate methods for discriminating any two quantum states. One geometric method on the basis of Bures angle was introduced to estimate the time \cite{Deffner13}. This method can provide a tight bound to the speed of the amplitude-damping decoherence. But it is just applicable to the evolution where the initial state is a pure one. To further study the open dynamics with respect to any mixed initial state, we have put forward another measure on the basis of trace distance. According to our previous results \cite{Hao15}, the validity of this measure have been demonstrated. Under some special conditions, this measure can provide an optimal time bound. For any initial mixed state $\hat{\rho}_0$, the trace distance in the evolution is expressed as $D_{\hat{\rho}}(t,0)=1-\frac{1}{4}\left \|\hat{\rho}_{t}-\hat{\rho}_{0}\right \|_1^2$. The evolved state $\hat{\rho}_{t}$ is given by a general dynamical map, i.e., $\dot{\hat{\rho}}_t =\hat{L}_t(\hat{\rho}_t)$ where $\hat{L}_t$ is a super operator with respect to the evolved state. Similar to the derivation of the Bures-angle approach, the ML-type bound and MT-type bound by trace distance are written as
\begin{equation}
\label{eq:13}
\tau_{QSL}=\max\{\frac{1}{\Lambda_{\tau_D}^1},\frac{1}{\Lambda_{\tau_D}^2},\frac{1}{\Lambda_{\tau_D}^{\infty}}\}2|1-D_{\hat{\rho}}(\tau_D,0)|,
\end{equation}
where $\Lambda_{\tau_D}^1=\frac{1}{\tau_D}\int_{0}^{\tau_D}\left \|\rho_t-\rho_0\right \|_1\cdot\left \|L_t(\rho_t)\right \|_1\,\mathrm{d}t$,
$\Lambda_{\tau_D}^2=\frac{\sqrt{n}}{\tau_D}\int_{0}^{\tau_D}\left \|\rho_t-\rho_0\right \|_1\cdot\left \|L_t(\rho_t)\right \|_2\,dt$ and
$\Lambda_{\tau_D}^{\infty}=\frac{n}{\tau_D}\int_{0}^{\tau_D}\left \|\rho_t-\rho_0\right \|_1\cdot\left \|L_t(\rho_t)\right \|_{\infty}\,\mathrm{d}t$. The Schatten $p$ norm $||A||_p=\left[ \sum_{i}\lambda_i^p \right]^{1/p}$ where $\lambda_i$ are the singular values of the operator $A$ in the descending sequence and $\lambda_1$ is the maximal singular value.

To investigate the $\mathrm{QSL}$ time for single-qubit thermal decoherence, we simply select the dressed state $|1\rangle$ as an initial state,
By use of Eq. (11), the evolved state $\hat{\bar{\rho}}_{t}$ can be obtained as
\begin{equation}
\label{eq:14}
\bar{\rho}_t=\frac{1}{2}\begin{pmatrix}
                   1+e^{-p(t)}[1+\delta(t)] & 0 \\
                   0 & 1-e^{-p(t)}[1+\delta(t)]
                  \end{pmatrix}.
\end{equation}
The evolution from initial state $\hat{\bar{\rho}}_0$ to final state $\hat{\bar{\rho}}_{\tau_D}$ by a given driving time $\tau_D$ is examined.
$\gamma_0/\lambda=0.1$ is assumed in the weak coupling regime and the driving time $\tau_D=2$.
Figure $1$ demonstrates the value of $\tau_{QSL}/\tau_D$ as a function of temperature $T$ of the
structured reservoir and the driving strength $\Omega$ of laser field. When the driving strength is small enough, i.e., $\Omega<2$, the $\tau_{QSL}/\tau_D$ exhibits a plateau called no speed-up region and its value is equal to $1$. But the value $\tau_{QSL}/\tau_D$ decreases with increasing the temperature in the strong-driving case. In other words, the increase of the temperature may contribute to the occurrence of the potential capacity for speedup evolution. The result can be understood by the fact that the environmental temperature can accelerate the mixing process of quantum states. The oscillation of $\tau_{QSL}/\tau_D$ between $0$ and $1$ is also seen with enlarging the driving strength of the classical field. However, when $\Omega>13\lambda$, the impacts of the driving field on $\tau_{QSL}/\tau_D$ are trivial.

The behavior that a quantum system can speed up in the process of dynamics can be explained by the relational expression between
$\tau_{QSL}/\tau_D$ and the population of the excited state \cite{Deffner13,Xu14,Zhang14},
\begin{equation}
\label{eq:15}
\frac{\tau_{QSL}}{\tau_D}=\frac{1}{2}\cdot \frac{(P_{\tau_D}-1)^2 }{\int_{0}^{\tau_D}|(P_t-1)\dot{P_t}| dt},
\end{equation}
where $P_t=\langle 1|\bar{\rho}_t|1 \rangle$ denotes the excited-state population. The above equation shows the value $\tau_{QSL}/\tau_D$ depends on the rate of the excited population $\dot{P}_t$. When $\dot{P}_t<0$ in the whole driving time, i.e. $|\dot{P_t}|=-\dot{P_t}$, the value $\tau_{QSL}/\tau_D=1$ that represents no speed-up process. While $\dot{P}_t>0$ at some times, the value $\tau_{QSL}/\tau_D<1$ that represents that the evolution may have the potential capacity for speedup. It is clearly seen from Figure 2 that the temperature of the reservoir and driving strength of laser field have impact on $\dot{P}_t$. If there exists the increase of the excited population during a given driving time, the quantum evolution could be of speed-up potential capacity.

It is also necessary to study the $\mathrm{QSL}$ time for open multi-particle system. We consider the open quantum system evolution starting from the $N-$qubit state $|\bar{\psi}^{(N)}\rangle=\frac {1}{\sqrt{2}}(\prod_{k=1}^{N}|1\rangle_k+\prod_{k=1}^{N}|0\rangle_k)$. This state is referred to as a maximally entangled state. For the simplest case of $N=2$, the expression of $\tau_{QSL}/\tau_D$ takes the form
\begin{equation}
\label{eq:16}
\frac{\tau_{QSL}}{\tau_D}=\frac{1}{2}\cdot \frac{[a_{\tau_D}(1-a_{\tau_D})+c_{\tau_D}(1-c_{\tau_D})+X_{\tau_D}]^2}
{\int_{0}^{\tau_D}[a(1-a)+c(1-c)+X][|\dot{a}(1-2a)+\dot{c}(1-2c)|+Y] dt},
\end{equation}
where $X=\frac{1}{2}\{(a^2+c^2-1)^2+[(1-a)^2+(1-c)^2-1]^2+2(b^2-1)({b^{\ast}}^2-1)+2|(a^2+c^2-1)[(1-a)^2+(1-c)^2-1]-(b^2-1)({b^{\ast}}^2-1)|\}^{\frac{1}{2}}$
, $Y=[(a\dot{a}+c\dot{c})^2+(a\dot{a}+c\dot{c}-\dot{a}-\dot{c})^2+2|b\dot{b}|^2+2|(a\dot{a}+c\dot{c})(a\dot{a}+c\dot{c}-\dot{a}-
\dot{c})-|b\dot{b}|^2|]^{\frac{1}{2}}$ and the notation $\dot{S}=\frac{\partial S}{\partial t},(S=a,b,c)$.

The value $\tau_{QSL}/\tau_D$ as a function of parameter $T$ for initial state $\frac{1}{\sqrt{2}}(|11\rangle+|00\rangle)$ is plotted in Figure 3.
To further demonstrate the effects of entanglement on $\mathrm{QSL}$ time, we take into account the circumstances of $0.1\leq\Omega\leq0.3$ and $0\leq{T}\leq0.8$ where there is no potential capacity for speed-up evolution of single system. In the case of the weak driving strength $\Omega=0.1$ at zero-temperature, the value $\tau_{QSL}/\tau_D<1$ due to entanglement \cite{Frowis12,Liu15} of initial state. Figure 3 shows that the evolution of the two-qubits system emerges gradual deceleration process for $T>0$ under different driving strength $\Omega$. Besides it, the increase of parameter $\Omega$ plays a positive role in the reduction of the value $\tau_{QSL}/\tau_D$. Therefore, the method of controlling parameters $T$ and $\Omega$ is also feasible to accelerate quantum evolution for two entangled systems.

Furthermore, we investigate the scaling property of $\mathrm{QSL}$ time for open $N-$qubit noninteracting entangled system. In Figure 4, the horizontal axis represents the amount of qubits and the corresponding values $\tau_{QSL}/\tau_D$ are marked by dots. The value $\tau_{QSL}/\tau_D$ first exhibits decay with the number and then decrease in tiny range. It is found out that the $\tau_{QSL}/\tau_D$ tends to be stable for the entangled system with $N>7$. This scaling property shows that the entanglement of the open system can only improve the potential capacity for quantum speedup to a certain extent.

To further demonstrate the validity of the Lorentzian spectral density approach in the weak-coupling limit, we treat another equivalent dissipation model \cite{Thorwart04,Goorden04,Zueco08}. An effective two-level system can be weakly coupled to the normal cavity mode with the frequency $\omega_0$ and operator $\hat{a}$. $g$ denotes the weak interaction strength between the two-level system and the normal mode. The new bath is restricted to the remaining oscillator modes coupled to the normal mode and the Hamiltonian can be written as
\begin{equation}
\label{eq:17}
\hat{H}=\hat{H}_S+(\hat{a}^{\dag}+\hat{a})\sum_j\nu_j(\hat{b}_j^{\prime\dag}+\hat{b}^{\prime}_j)+\sum_j\omega^{\prime}_j\hat{b}_j^{\prime\dag}\hat{b}^{\prime}_j,
\end{equation}
where $\hat{H}_S=\frac{1}{2}[(\omega_0-\omega_L)\hat{\sigma}_z+\Omega\hat{\sigma}_x]+g(\hat{a} \hat{\sigma}_{+}+\hat{a}^{\dag}\hat{\sigma}_{-})+\omega_0\hat{a}^{\dag}\hat{a}+(\hat{a}^{\dag}+\hat{a})^2\sum_j\nu_j^2/2{\omega}^{\prime}_j$.
The bath is described by the Ohmic spectral density function $J_{Ohm}(\omega^{\prime})=\sum_j|\nu_j|^2\delta(\omega^{\prime}-{\omega}^{\prime}_j)=\gamma\omega^{\prime}\exp(-\omega^{\prime}/\omega_c)$ where the decay rate $\gamma$ is small and the cutoff frequency is $\omega_c$. The effective spectral density function $J_{eff}(\omega^{\prime})=\frac {2\alpha\omega^{\prime}\omega_0^4}{(\omega_0^2-\omega^{\prime 2})^2+(2\pi\gamma\omega^{\prime}\omega_0)^2}$ is considered in this case. The
parameter $g$ and weak coupling strength $\alpha$ follows as $\alpha=8\gamma\frac {g^2}{\omega_0}$. The condition $g\ll \gamma$ is
feasible in the weak coupling approximation according to Refs. \cite{Thorwart04,Goorden04}. Similar to the result gained by the Lorentzian spectral density in Figure $1$, Figure $5$ also shows that the control of the strength of the classical field or temperature is useful for the acceleration of quantum evolution in the new dissipation model.

\section{Conclusions}

An external classical field is applied to the $N-$particle damped Jaynes-Cummings model at low temperatures when the system-environment coupling is weak. By means of time-convolutionless master equation and the technique of projection operator, we obtain the dynamical process of the whole open system in the weak coupling regime. In spite of weak couplings, the quantum evolution may take on the potential capacity for speedup by the control of the driving strength in finite-temperature environments. For the multi-qubit open system, we demonstrate that multi-qubit entanglement of open system can be thought of as resources for accelerating quantum decoherence in thermal environment. The manipulation of classical laser field can provide us a new way to accelerate the quantum evolution.

\newpage
\begin{figure}
  \centering
  \includegraphics[width=0.5\textwidth]{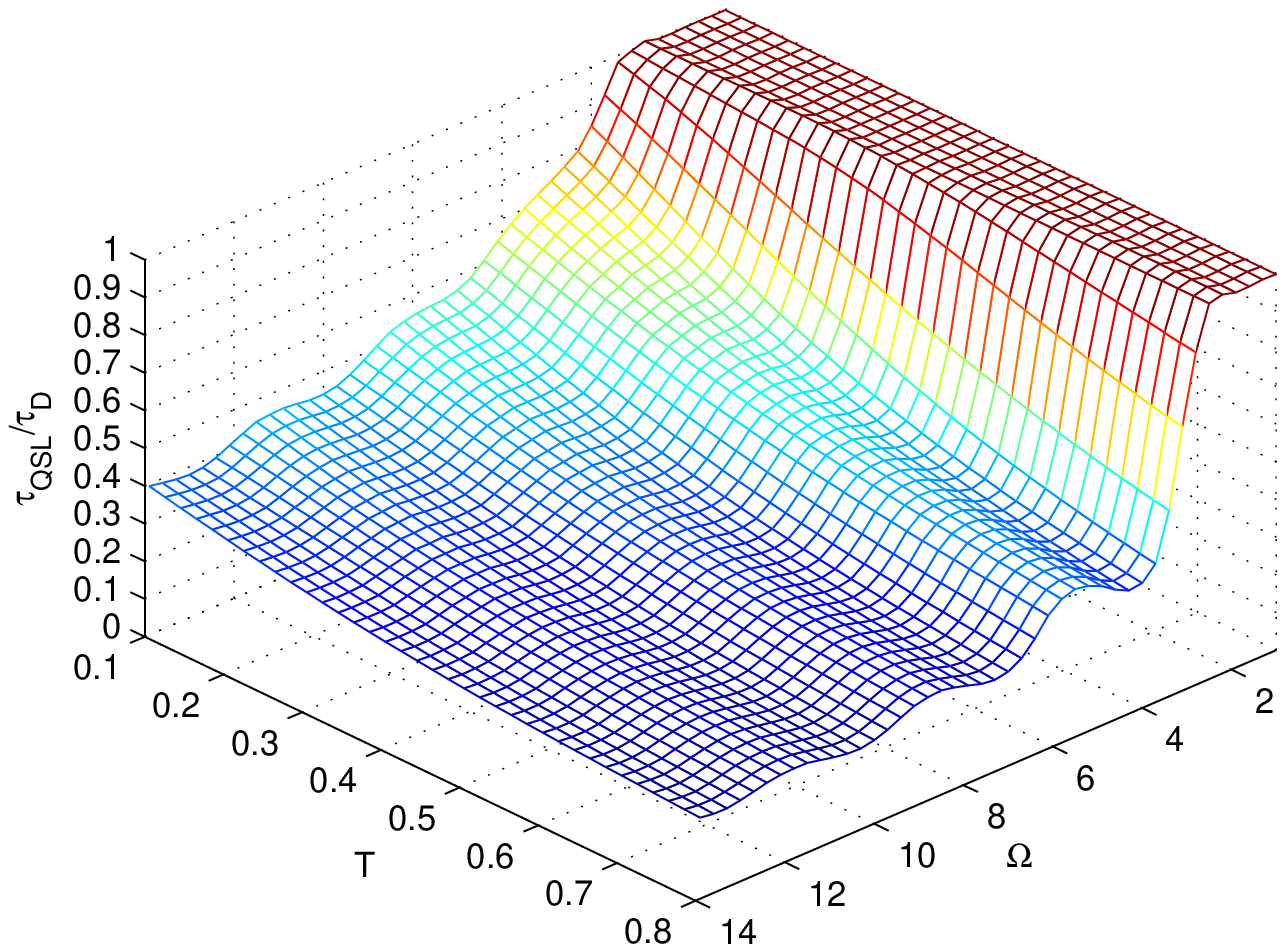}\\
  \caption{$\tau_{QSL}/\tau_D$ is plotted as a function of temperature $T$ of the structured reservoir and the driving strength
  $\Omega$ of laser field for the initial excited state.
  The parameters used here are $\lambda=1$, $\omega_0=1.1$ and $\omega_L=1$.}\label{1}
\end{figure}

\begin{figure}
  \centering
  \includegraphics[width=0.5\textwidth]{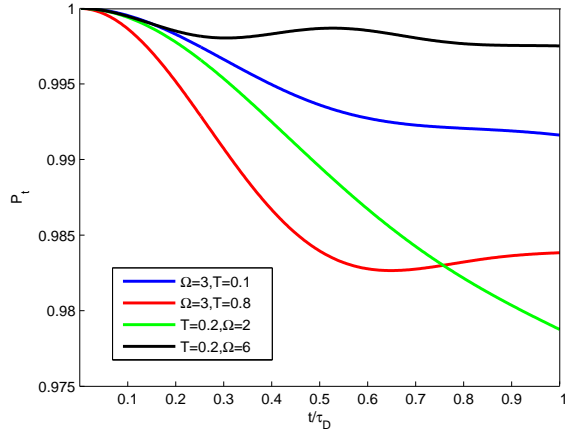}\\
  \caption{The population of the excited state as a function of the evolved time $t$ is plotted in the case of
  the different choice of temperature $T$ and the driving strength $\Omega$. Parameters are $\lambda=1$,
  $\omega_0=1.1$ and $\omega_L=1$.}\label{2}
\end{figure}

\begin{figure}
  \centering
  \includegraphics[width=0.5\textwidth]{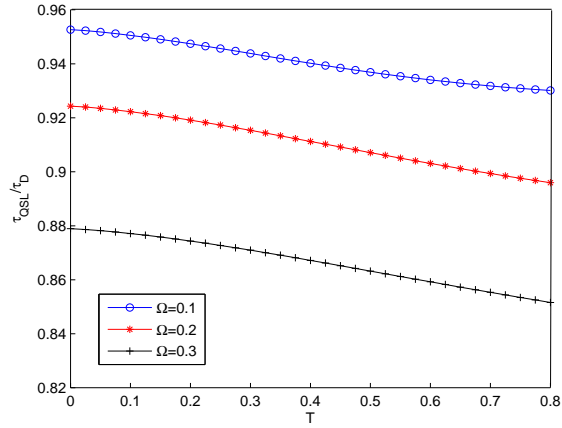}\\
  \caption{The value $\tau_{QSL}/\tau_D$ for the initial state $\frac{1}{\sqrt{2}}(|11\rangle+|00\rangle)$ as
  a function of temperature of reservoir. The three lines respectively denote
  the condition of $\Omega=0.1, 0.2, 0.3$. Parameters are $\lambda=1$,
  $\omega_0=1.1$ and $\omega_L=1$.}\label{3}
\end{figure}

\begin{figure}
  \centering
  \includegraphics[width=0.5\textwidth]{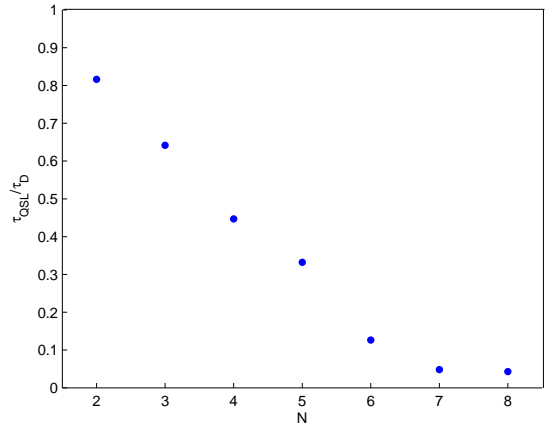}\\
  \caption{The value $\tau_{QSL}/\tau_D$ as a function of the number of qubits
  for multiqubit systems driven by laser field at finite temperature $(T=0.1)$.
  Parameters are $\Omega=0.4$, $\lambda=1$, $\omega_0=1.1$ and $\omega_L=1$.}\label{4}
\end{figure}

\begin{figure}
  \centering
  \includegraphics[width=0.5\textwidth]{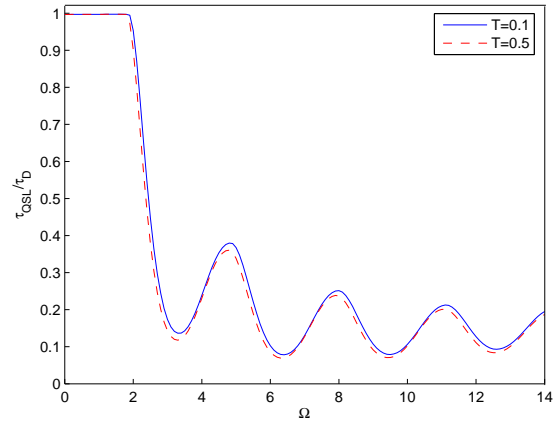}\\
  \caption{The value $\tau_{QSL}/\tau_D$ as a function of the driving strength $\Omega$
  is plotted respectively at finite temperature $T=0.1$ and $T=0.5$ for
  the new dissipation model with $\tau_D=2$, $\gamma=0.1$, $g=0.01$,
   $\omega_0=1.1$ and $\omega_L=1$.}\label{5}
\end{figure}

\end{document}